\documentstyle[12pt]{article}
\normalbaselines
\begin{document}
\bibliographystyle{unsrt}

\vbox {\vspace{6mm}}

\begin{center}

{\large \bf MINIMUM UNCERTAINTY AND SQUEEZING IN DIFFUSION \\
            PROCESSES AND STOCHASTIC QUANTIZATION\footnote{Proceedings
            of the Third International Workshop on
            Squeezed States and Uncertainty Relations, held in
            Baltimore, 10-13 August 1993 (NASA Conference Publication,
            in print).}} \\ [7mm]
S. De Martino, S. De Siena, F. Illuminati, and G. Vitiello \\ [5mm]
{\it Dipartimento di Fisica, Universit\`{a} di Salerno, and \\
     INFN Sezione di Napoli, 84081 Baronissi (Salerno), Italia} \\ [5mm]
\end{center}

\vspace{2mm}

\begin{abstract}
We show that uncertainty relations, as well as
minimum uncertainty coherent and squeezed states,
are structural properties for diffusion processes.
Through Nelson stochastic quantization
we derive the stochastic image of
the quantum mechanical coherent and squeezed states.
\end{abstract}

\section{Introduction}

It is well known that the theory of stochastic processes is
a powerful tool in the study of the interplay
between probabilistic and deterministic evolution \cite{gard85}.
In quantum mechanics, and in particular in quantum optics,
such interplay is expressed by the states of minimum uncertainty,
the coherent \cite{klaud85} and squeezed states \cite{stoler70},
which are viewed as the ``most classical" states.

In this paper we report on a recent derivation \cite{demar93} of
uncertainty relations
for classical stochastic processes of the diffusion type, and
we determine the diffusion processes
of minimum uncertainty (MUDPs). We find that
a special class among them is associated
to Gaussian probability distributions with time-conserved covariance
and mean value with classical time evolution:
we refer to them as strictly coherent MUDPs.
We will also identify Gaussian MUDPs with time-dependent covariance
and conserved expectation value: we refer to them as broadly coherent
MUDPs.
By exploiting Nelson's stochastic
quantization scheme \cite{guerra81}, we will show that the strictly
coherent MUDPs
provide the stochastic image of the standard quantum mechanical
coherent and squeezed coherent states, while the broadly coherent
MUDPs are associated with the phenomenon of time-dependent
squeezing.

Our study is motivated
by the possibility that the formalism of stochastic processes
offers to treat on the same footing, in a unified mathematical
language, the interplay between
fluctuations of different nature, for instance quantum and
thermal \cite{zann82}.

Beyond the case of diffusion processes, it is interesting to note
that coherence and squeezing have recently
emerged in other contexts wider than
quantum mechanics (\cite{sud93},
\cite{vit93}).

\section{Uncertainty and Coherence in Diffusion Processes}
In what follows, without lack of generality,
we will consider a one-dimensional random
variable $q$.
The associated diffusion process
$q(t)$ obeys
\begin{equation}
dq(t) = v_{(+)}(q(t),t)dt + {\nu}^{1/2}(q(t),t)dw(t) \, ,
\; \; \; dt > 0 \, ,
\end{equation}

\noindent where $v_{(+)}(q(t),t)$, is the forward drift,
$\nu (q(t),t)$ is the diffusion coefficient, and $dw(t)$ is a
Gaussian white noise, superimposed on the otherwise deterministic
evolution, with expectation $E(dw(t)) =0$ and covariance
$E(dw^{2}(t)) =2dt$.
The forward and the backward drifts $v_{(+)}(x,t)$ and
$v_{(-)}(x,t)$ are defined as
\begin{eqnarray}
v_{(+)}(x,t) & = & \lim_{\Delta t \rightarrow 0^{+}}E
\left( \frac{q(t + \Delta t) - q(t)}{\Delta t}\left.
\right| q(t) = x \right) \, , \nonumber \\
&  & \\
v_{(-)}(x,t) & = & \lim_{\Delta t \rightarrow 0^{+}}E
\left( \frac{q(t) - q(t - \Delta t)}{\Delta t}\left. \right|
q(t) = x \right) \, . \nonumber
\end{eqnarray}

The definitions of $v_{(+)}$ and $v_{(-)}$ are not independent, but
related by \cite{guerra81}
\begin{equation}
v_{(-)}(x,t) = v_{(+)}(x,t) -
\frac{2\partial_{x}(\nu(x,t)\rho(x,t))}{\rho(x,t)} \, .
\end{equation}

It is now convenient to define the osmotic velocity $u(x,t)$ and
the current velocity $v(x,t)$
\begin{eqnarray}
u(x,t) & = & \frac{v_{(+)}(x,t) - v_{(-)}(x,t)}{2} =
\frac{\partial_{x}(\nu(x,t)\rho(x,t))}{\rho(x,t)} \, , \nonumber \\
& & \\
v(x,t) & = & \frac{v_{(+)}(x,t) + v_{(-)}(x,t)}{2} \, . \nonumber
\end{eqnarray}

{}From the former definitions it is clear that $u(x,t)$ ``measures" the
non-differentiability of the random trajectories, controlling the
degree of stochasticity.
In the deterministic limit $u$ vanishes, and $v(x,t)$ goes to the
classical velocity $v(t)$.

Finally, we have the continuity equation
\begin{equation}
\partial_{t}\rho(x,t) = -\partial_{x}(\rho(x,t)v(x,t)) \, .
\end{equation}

It is straightforward to check that $E(v_{(+)}) = E(v_{(-)})
= E(v)$, and $E(u)=0$. Further,
\begin{equation}
E(v) = \frac{d}{dt}E(q) \; \; \; \; \; \; \; \forall t \, .
\end{equation}

For the product $qu$,
we have $|E(qu(q,t))| = E(\nu(q,t))$.
By Schwartz's inequality, the r.m.s.
deviations $\Delta q$ and $\Delta u$ satisfy
\begin{equation}
\Delta q \Delta u \geq E(\nu (q,t)).
\end{equation}

Inequality (7) is the uncertainty relation
for any diffusion process. Equality in (7)
defines the MUDPs. Saturation
of Schwartz's inequality yields $u(x,t)=C(t)(x-E(q))$,
where $C(t)$ is an arbitrary function of time.
Considering constant $\nu$ and time-dependent $\nu$,
in both cases we
obtain a Gaussian minimum uncertainty density:
\begin{equation}
\rho(x,t) = \frac{1}{\sqrt{2\pi(\Delta q)^{2}}}
\exp{\left[-\frac{(x - E(q))^{2}}{2(\Delta q)^{2}}\right]} \, ,
\end{equation}

\noindent where $2(\Delta q)^{2}=-\nu(t)/C(t)$.

{}From eq. (5) we can determine the current
velocity:
\begin{equation}
v(x,t) = \frac{d}{dt}E(q) + \frac{1}{\Delta q}
\left( \frac{d}{dt}\Delta q\right)
F(x,t) \, ,
\end{equation}

\noindent where
\begin{equation}
F(x,t) = x - E(q) + E(q)\exp{\left[ \frac{x^{2} -
2xE(q)}{2(\Delta q)^{2}}\right]}
\, .
\end{equation}

Eqs. (8)-(10) lead to the stochastic differential equation
obeyed by any MUDP:
\begin{equation}
dq(t) = [A(t) + B(t)q(t)]dt + {\nu}^{1/2}(t)dw(t) \, .
\end{equation}

It is interesting to observe that (11) defines the so-called
{\it linear processes in narrow sense}. When $A(t)=0$
they are the time-dependent Ornstein-Uhlenbeck processes. These
last ones play a natural role in the theory of low noise systems
[1], which are thus found to be related with MUDPs.

The possible choices of $E(q)$ and $\Delta q$ in (8) are
not independent: taking the expectation value of in (9)-(10),
and reminding (6) one has that either
\begin{equation}
\left\{ \begin{array}{ll}
\Delta q  & =  \mbox{const.} \; \; \; \; \; \; \forall t \, , \\
E(q)  & = j(t) \, , \end{array}
\right.
\end{equation}

\noindent or
\begin{equation}
\left\{ \begin{array}{ll}
\Delta q & =  k(t) \, , \\
E(q) & =  0 \; \; \; \; \; \; \forall t \, , \end{array}
\right.
\end{equation}

\noindent where $j(t)$ and $k(t)$ are arbitrary functions
of time, and we have chosen for simplicity $q(t=0)=0$.
Consider first case (12): $\Delta q$
does not spread; also, it is immediate to verify that
the expectation value of the process $E(q)$ follows a classical
trajectory:
\begin{equation}
v(x,t)=\frac{d}{dt}E(q)=v(t) \, ,
\end{equation}

As a consequence, MUDPs of the
form (8) obeying (12) and (14) are
coherent in a sense precisely
analogous to that of quantum mechanical coherent
states: we will refer to them as strictly coherent MUDPs
and to processes (8) obeying (13) as broadly coherent MUDPs.

It is possible to discriminate on physical grounds the strictly
coherent MUDPs from the broadly coherent ones by observing
that (12) and (14) come as immediate consequence on imposing the
Ehrenfest condition
\begin{equation}
v(E(q),t)=\frac{d}{dt}E(q) \, ,
\end{equation}

\noindent so that the strictly coherent MUDPs can be viewed
as the most deterministic semi-classical processes.

Consider the scale transformation $x \rightarrow
e^{-s}x$ which authomatically implies $u \rightarrow e^{s}u$,
where $s$ is the scale parameter.
The Gaussian distribution (8) is form-invariant under this
transformation, while the uncertainty product (7) is strictly
invariant with $\Delta q \rightarrow
e^{-s}\Delta q$ and $\Delta u \rightarrow e^{s}\Delta u$.
We will show next that in the framework of Nelson
stochastic quantization
this transformation is just the squeezing
transformation of quantum mechanics.
In this context, broadly coherent MUDPs
are of special interest when considering
time-dependent squeezing.

\section{Nelson Diffusions}
A very important class of diffusion
processes (Nelson diffusions) in physics has been introduced
by Nelson in his stochastic formulation of quantum
mechanics \cite{guerra81}.

To each single-particle quantum state
$\Psi = \exp{ \left[ R + \frac{i}{\hbar}S\right] }$,
Nelson stochastic quantization associates the diffusion process
$q(t)$ with
\begin{equation}
\nu=\frac{\hbar}{2m} \, , \; \; \; \; \; \; \; \; \;
\rho(x,t)=|\Psi(x,t)|^{2}\, ,  \; \; \; \; \; \; \; \; \;
v(x,t) = \frac{1}{m}\frac{\partial S(x,t)}{\partial x} \, ,
\end{equation}

\noindent where $m$ is the mass of the particle. At the dynamical
level, the Schroedinger equation with potential $V(x,t)$
is equivalent to the
Hamilton-Jacobi-Madelung equation
\begin{equation}
{\partial}_{t}S(x,t) + \frac{({\partial}_{x} S(x,t))^{2}}{2m}
- \frac{\hbar^{2}}{2m}\frac{\partial_{x}^{2}\rho^{1/2}(x,t)}
{\rho^{1/2}(x,t)} = -V(x,t) \, .
\end{equation}

It is well known \cite{defalco82} that for Nelson diffusions
the uncertainties $\Delta q$ and $\Delta u$ are related to
the quantum mechanical uncertainties $\Delta \hat{q}$
and $\Delta \hat{p}$ of the position and momentum operators
$\hat{q}$ and $\hat{p}$ by
\[
\Delta \hat{q} =  \Delta q \, , \; \; \; \; \; \; \;
(\Delta \hat{p})^{2} = m[(\Delta u)^{2} + (\Delta v)^{2}] \, ,
\]
\begin{equation}
\end{equation}
\[
(\Delta \hat{q})^{2} (\Delta \hat{p})^{2} \geq
(\Delta q)^{2} (\Delta mu)^{2} \geq \frac{{\hbar}^{2}}{4} \, .
\]

Minimum uncertainty Nelson diffusions (MUNDs) are MUDPs.
Correspondingly, we will speak of strictly and broadly coherent
MUNDs. By solving (17) for MUNDs we
obtain $V(x,t)$ and the classical
equations of motion for $E(q)$.
For strictly coherent MUNDs (12) we have
\newpage
\[
V(x,t)  =  \frac{m}{2}\omega^{2}x^{2} + f(t)x +
V_{0}(t)\, ,\; \; \; \;
\; \; \; \; \; \omega^2 = \frac{\hbar^2}{4m^{2}(\Delta q)^4}
\; \; ,
\]
\begin{equation}
\end{equation}
\[
\frac{d^2}{dt^2}E(q) +  \omega^{2}E^{2}(q)
=  f(t)  \, .
\]

When the arbitrary constants $f(t)$ and $V_{0}(t)$ vanish,
eqs. (19) are those of the classical harmonic oscillator
and the associated quantum states
are the standard Glauber coherent states; when
$f(t)=$ const. we have the Klauder-Sudarshan displaced oscillator
coherent states; finally, when $f(t)$ is truly time-dependent,
we obtain
the Klauder-Sudarshan driven oscillator coherent states \cite{klaud70}.

For broadly coherent MUNDs we have instead
\begin{equation}
V(x,t) = \frac{1}{2}m\omega^{2}(t)x^2 + V_{0}(t) \, , \, \;
\; \; \; \; \;
\omega^{2}(t) = \dot{g}(t) + 2g^{2}(t) -
\frac{\hbar^{2}}{8m^{2}(\Delta q)^{4}} \, ,
\end{equation}

\noindent where $g(t) = (\Delta q)^{-1}d\Delta q/dt$ must be such that
$\omega^{2}(t)$ is positive. Eq. (20) describes the parametric
oscillator potential, associated to the feature of
time-dependent squeezing.

Furthermore, we can identify among MUNDs
those corresponding either to Heisenberg or to Schroedinger
minimum uncertainty. The key relation, easy to prove, is
\begin{equation}
E(vq) - E(v)E(q) = \frac{
<\{ \hat{Q} , \hat{P} \}>_{\psi}}{2} \, ; \; \;
\; \; \; \hat{Q}=\hat{q} -
<\hat{q}>_{\psi} \, , \; \; \hat{P}=
\frac{\hat{p} - <\hat{p}>_{\psi}}{m} \, ,
\end{equation}

\noindent where $<\{\cdot , \cdot \}>_{\psi}$ denotes the expectation
of the anti-commutator in the state $\Psi$,
i.e. the Schroedinger part of the quantum mechanical
uncertainty.

Eqs. (18) and (21) show that the strictly coherent MUNDs (19)
exhaust the Heisenberg minimum uncertainty states, while the broadly
coherent MUNDs (20) form a subset of the Schroedinger minimum
uncertainty states.

Finally, we investigate the possibility of letting $\nu$ be
time-dependent in the context of quantum mechanics. From
the first of equations (16) this means
letting either $m$ or $\hbar$ be functions of time.

This latter case
seems a bit speculative at this stage.
We thus fix our attention on the case of time-dependent mass $m(t)$
and constant $\hbar$.

For such systems it can be immediately verified that the
Nelson scheme (16)-(17)
still holds with $m(t)$ replacing $m$.
Considering the most interesting case of strictly coherent MUNDs,
which means choosing $C(t) \propto \nu(t)$, and solving
(17) we obtain
\[
V(x,t) =  \frac{1}{2}m(t)\omega^{2}(t)x^{2} + f(t)x + V_{0}(t) \, ,
\; \; \; \; \; \; \;
\omega^{2}(t) = \frac{\hbar^{2}}{4m^{2}(t)(\Delta x)^{4}} \, ,
\]
\begin{equation}
\end{equation}
\[
\frac{d^{2}}{dt^{2}}E(q) +
\frac{\dot{m}(t)}{m(t)}\frac{d}{dt}E(q) +
\omega^{2}(t)E(q) = \frac{f(t)}{m(t)} \, ,
\]

\noindent where $f(t)$, $V_{0}(t)$ are arbitrary functions of time.
Eqs. (22) supplemented with $m(t) = m_{0}e^{\Gamma(t)}$
define the dynamics of the damped
parametric oscillator.
The stochastic approach thus sheds new light in a
unified treatment on the study of
quantum dissipative oscillators,
for it allows to derive for the expectation value
the dynamical equation (22)
that was so far unknown.

In conclusion, we have shown that the quantum mechanical concepts
of uncertainty, coherence, and squeezing can be imported
in the probabilistic arena of diffusion processes.
This appears to be possible because of a subtle interplay
between fluctuations, control, and optimization.
Conversely, we may also say that these features of quantum
mechanics can be traced back and related to
general properties of diffusion processes.

Work on this subject is in progress, and includes application of
our scheme
to polymer dynamics and chemical reactions, uncertainty relations
in field theory and dynamical systems on lattices and manifolds.

\end{document}